\documentclass[amsmath,amssymb,nofootinbib,superscriptaddress,showpacs,showkeys]{revtex4-2}
\usepackage{graphicx}
\usepackage{dcolumn}
\usepackage{bm}
\usepackage{rotating} 
\usepackage{subfigure}
\usepackage{physics}
 \usepackage{multirow} 
 \usepackage[table]{xcolor}
\usepackage[colorlinks=true, citecolor=blue, urlcolor = magenta, linkcolor= red, bookmarks=true]{hyperref}
\usepackage{orcidlink}

\begin{document}

\title{Lyapunov Exponent Approach to Phase Structure of Schwarzschild AdS Black Holes Surrounded by a Cloud of Strings}

\author{Arun Kumar \texorpdfstring{\href{https://orcid.org/0000-0001-8461-5368}{\orcidlink{0000-0001-8461-5368}}{}}} \email{arunbidhan@gmail.com} 
\affiliation{Institute for Theoretical Physics and Cosmology, Zhejiang University of Technology, Hangzhou 310023, China}
\author{Qiang Wu\texorpdfstring{\href{https://orcid.org/0000-0002-5483-4903}{\orcidlink{0000-0002-5483-4903}}{}}} \email{wuq@zjut.edu.cn} 
\affiliation{Institute for Theoretical Physics and Cosmology, Zhejiang University of Technology, Hangzhou 310023, China}
\author{Tao Zhu\texorpdfstring{\href{https://orcid.org/0000-0003-2286-9009}{\orcidlink{0000-0003-2286-9009}}{}}} \email{zhut05@zjut.edu.cn} 
\affiliation{Institute for Theoretical Physics and Cosmology, Zhejiang University of Technology, Hangzhou 310023, China}
\author{Sushant~G.~Ghosh \texorpdfstring{\href{https://orcid.org/0000-0002-0835-3690}{\orcidlink{0000-0002-0835-3690}}{}}}\email{sghosh2@jmi.ac.in}
\affiliation{Centre for Theoretical Physics, 
Jamia Millia Islamia, New Delhi 110025, India}
\affiliation{Astrophysics and Cosmology Research Unit, 
School of Mathematics, Statistics and Computer Science, University of KwaZulu-Natal, Private Bag 54001, Durban 4000, South Africa}

\begin{abstract}

We investigate Schwarzschild black holes in anti-de Sitter (AdS) spacetimes surrounded by a cloud of strings (BH-AdS-CoS), incorporating both electric- and magnetic-like components of the string bi-vector. Thermodynamically, these systems exhibit small/intermediate/large black hole phases with first- and second-order transitions governed by the string parameter $c_0$. Dynamically, we probe the phase structure using Lyapunov exponents $\lambda$ from unstable circular geodesics. For massless particles ($\delta = 0$), analytical expressions $\lambda$ reveal multivalued behavior in first-order transition regimes ($c_0 < c_{\text{cri}}$), with branches mapping to thermodynamic phases ($\lambda_{\text{SBH}}, \lambda_{\text{IBH}}, \lambda_{\text{LBH}}$). The discontinuity $\Delta\lambda = \lambda_{\text{SBH}} - \lambda_{\text{LBH}}$ at $T_p$ follows mean-field scaling: $\Delta\lambda / \lambda_{\text{cri}} \propto (T_\text{cri} - T)^{1/2} \quad (\beta = 1/2)$. For massive particles ($\delta = 1$), numerical computation of timelike geodesics confirms $\lambda$ as an order parameter, with critical exponent $\beta = 1/2$ universally. Key distinctions emerge: $\lambda\to 1$ asymptotically for photons, while $\lambda\to 0$ in the significant black hole phase for massive particles due to vanishing unstable orbits. The transition of $\lambda$ from multivalued to single-valued at $c_0 = c_{\text{cri}}$ establishes it as a universal dynamical probe of black hole criticality. The universal critical exponent of 1/2 for \(\Delta\lambda\) further reinforces the analogy with conventional thermodynamic systems.  Our results confirm a direct connection between the thermodynamic phase structure of BH-AdS-CoS and the dynamics of test particles, with the Lyapunov exponent emerging as a sensitive diagnostic of black hole criticality.

\end{abstract}
\maketitle

\section{Introduction}

The thermodynamics of black holes (BHs), once regarded as a conceptual novelty, has become a central component of gravitational physics. The discovery that BHs obey laws analogous to those of thermodynamics \cite{Bekenstein:1973ur, Hawking:1974sw}—and radiate with a temperature proportional to their surface gravity—has deepened our understanding of gravity, quantum field theory, and statistical mechanics. The pioneering work of Hawking and Page revealed a phase transition between thermal AdS space and Schwarzschild-AdS BHs, now famously known as the Hawking-Page transition \cite{Hawking:1982dh}. A pivotal development identified the Hawking-Page phase transition in asymptotically anti-de Sitter (AdS) spacetimes, corresponding to a transition between thermal radiation and stable large AdS BHs \cite{Hawking:1982dh}.

The phase structure of AdS BHs has since been extensively explored in extended phase space thermodynamics, where the negative cosmological constant is treated as a thermodynamic pressure \cite{Kubiznak:2012wp}. This analogy reveals that charged and rotating AdS BHs exhibit a rich variety of phase behaviours similar to those seen in van der Waals fluids \cite{Chamblin:1999tk, Chamblin:1999hg}, including first- and second-order phase transitions, reentrant phase transitions \cite{Gunasekaran:2012dq, Altamirano:2013ane}, triple points \cite{Altamirano:2013uqa}, and critical phenomena characterized by universal exponents. The phase structure of AdS BHs has also been analysed through the null geodesics parameters, namely photon orbit radius and the critical impact parameter, in the literature \cite{Wei:2018aqm, Wei:2017mwc, NaveenaKumara:2019nnt, Zhang:2019tzi, Kumar:2024sdg, Sood:2024ufi}.

An exciting class of models involves BHs coupled to matter fields, such as scalar fields or exotic matter distributions. One such example is the BH surrounded by a cloud of strings (CoS). Letelier established its formalism as a classical matter source derived from the Nambu-Goto action \cite{Letelier:1979ej}. In this model, the matter source is described by the Nambu-Goto action averaged over a statistical ensemble of strings. The energy-momentum tensor derived from such configurations leads to non-trivial modifications to the spacetime geometry. When embedded in an AdS background, these string cloud BHs (CoS-AdS) show thermodynamic behaviour sensitive to the CoS parameters, adding a tunable structure to the phase space \cite{Singh:2020nwo, Herscovich:2010vr}. BHs surrounded by a CoS have been widely explored in various gravitational theories, including Lovelock gravity and its generalizations, revealing rich thermodynamic and dynamical structures \cite{Ghosh:2014dqa, Ghosh:2014pga, Lee:2014dha, Ganguly:2014cqa, Lee:2015xlp}. These studies have also demonstrated how CoS influence BH stability, horizon structure, and critical behaviour in classical and quantum-corrected settings \cite{Singh:2020nwo, Sood:2022fio, Kumar:2024qon, Kumar:2024bls}. BH thermodynamics reveals equilibrium properties, but studying the dynamics of test particles probes stability and near-horizon physics. In particular, the Lyapunov exponent $\lambda$—which characterises the instability timescale of circular geodesics—is emerging as a powerful diagnostic tool for probing dynamical and thermodynamic transitions in BH spacetimes \cite{Cardoso:2008bp}.

The Lyapunov exponent $\lambda$ governs the exponential divergence of nearby geodesics and has been related to the imaginary part of the eikonal limit quasinormal modes (QNMs) of BHs \cite{Konoplya:2017wot}. In chaotic systems, it provides a measure of dynamical instability. In BH physics, it has been argued to be bounded from above by $2\pi T$ (the so-called chaos bound) in holographic theories \cite{Maldacena:2015waa, Hashimoto:2016dfz}. This exponent has also been used for BHs to analyze time delay in strong lensing \cite{Wei:2013mda} and the stability of photon spheres \cite{Koga:2019uqd}.

The rate at which nearby trajectories in space diverge close to BHs is measured by the Lyapunov exponent. It has developed into a very useful tool for comprehending their stability and how their thermodynamics relate to it. Cornish and Levin \cite{Cornish:2003ig} and other early research emphasized the essential function of the Lyapunov exponent in assessing the instability of circular orbits.  They showed how to apply it to BH binaries, establishing a connection between geodesic instability and the predictability of relativistic trajectories.   This approach was subsequently extended to BHs in alternative gravity frameworks, such as the Kehagias-Sfetsos solution in Ho\v{r}ava-Lifshitz gravity \cite{Setare:2010zd}, and to standard solutions like Reissner–Nordstr\"{o}m, where the connection between Lyapunov exponents and quasinormal modes (QNMs) was clarified \cite{Pradhan:2012rkk}.

A surge of recent interest has further solidified the Lyapunov exponent's role as a diagnostic of both dynamical and thermodynamic features of BHs. In particular, its behavior has been used to identify the stability of null and timelike circular orbits and to characterize BH phase transitions in various spacetimes, including Hayward \cite{Mondal:2020uwp}, Kerr-Kiselev \cite{Mondal:2021exj}, BTZ-like models with nonlinear electrodynamics \cite{Giri:2021wxu}, and Einstein-Maxwell-Dilaton-Axion BHs \cite{Yu:2022tlr}. Giri and collaborators have also demonstrated its applicability in string-inspired and charged backgrounds \cite{Giri:2021kgc}. The Lyapunov exponent has been found to serve as a robust indicator of phase transitions, including those involving van der Waals-like behaviour, criticality, and second-order transitions, often exhibiting scaling behaviour and discontinuities across the critical temperature \cite{Guo:2022kio, Yang:2023hci, Kumara:2024obd, Shukla:2024tkw}. Notably, the universality and bounds of Lyapunov exponents have been explored across different BH types. For example, the existence of upper bounds in Kerr-Newman and de Sitter BHs has been derived via charged particle dynamics \cite{Kan:2021blg, Park:2023lfc}. The mathematical structure of these bounds and their geometric interpretations have been examined through detailed analyses of photon spheres and light rings \cite{Gallo:2024wju}. In AdS spacetimes, including Born-Infeld and Gauss-Bonnet BHs, Lyapunov exponents serve as efficient probes of BH thermodynamics, as demonstrated by studies like those conducted by Chen et al. \cite{Chen:2025xqc} and Lyu et al. \cite{Lyu:2023sih}. Furthermore, in order to further refine geodesic stability characterizations, Blaga et al.'s recent work \cite{Blaga:2023exi} focused on the Jacobi stability analysis in conjunction with Lyapunov exponents. 

According to recent studies, Lyapunov exponents' multivalued structure and discontinuity across critical points make them suitable as order parameters for BH phase transitions \cite{Gogoi:2024akv, Yang:2023hci, Guo:2022kio}. According to mean-field theory, van der Waals-like small-to-large AdS BH transitions have been observed in a number of models and show a finite jump $\Delta\lambda$ in the Lyapunov exponent at critical temperature that scales near criticality with exponent $\beta=1/2$ \cite{Zhao:2018kpz, Lyu:2023sih, Kumara:2024obd, Chen:2025xqc}. ~\citet{Letelier:1979ej} developed the BH-CoS, resulting metric function, $$f(r) = 1  - \frac{2M}{r}- a, $$ describes the string cloud's influence through the parameter \(a \), but its origin remains ad-hoc. Recently, ~\citet{Alencar:2025zyl} expanded this model by including a magnetic-like component, \(\Sigma_{23} \), and an electric-like component, \(\Sigma_{01} \). This work not only improves the original model. This opens new directions for exploring black hole thermodynamics and the observational consequences of string clouds ~\citet{Alencar:2025zyl}. In this study, we employ Lyapunov exponents to probe the thermodynamic phase structure of BH-AdS-CoS. By analysing the instability of circular geodesics—null and timelike—we demonstrate that the Lyapunov exponent serves as a universal dynamical order parameter, exhibiting multivalued behaviour and mean-field critical scaling near phase transitions.

This paper is structured as follows:  \textbf{Section~\ref{sec:action}} introduces the theoretical framework of a BH-AdS-CoS, including the action, metric, and derived thermodynamic properties. \textbf{Section~\ref{sec:particle}} probes the phase structure through Lyapunov exponents ($\lambda$) from unstable circular geodesics. The numerically calculated Lyapunov exponent $\lambda$ exhibits multivalued behaviour for massless particles, indicating first- and second-order phase transitions. We analyse effective potentials for massive particles, numerically compute $\lambda$ for timelike geodesics, and establish their role as order parameters while contrasting null and timelike behaviours.  
\textbf{Section~\ref{sec:conclusion}} summarises our results, emphasising the Lyapunov exponent's utility in diagnosing BH phase transitions and critical phenomena. 

\section{Black hole in AdS spacetime surrounded by clouds of strings}\label{sec:action}

The Einstein-Hilbert action coupled to a cosmological constant and a matter source describing a CoS is given by \cite{Letelier:1979ej, Zeldovich:1974uw}:
\begin{equation}\label{action}
S = \frac{1}{16\pi G}\int d^4x\sqrt{-g}\left[R - 2\Lambda\right] + S_M,
\end{equation}
where $R$ is the Ricci scalar curvature, $\Lambda$ is the negative cosmological constant related to the AdS radius $l$ through $\Lambda = -3/l^2$ \cite{Hawking:1982dh}, and $S_M$ represents the matter action. The matter content is described by the Nambu-Goto action, which effectively models a CoS as proposed in \cite{Letelier:1979ej}:
\begin{equation}\label{NambuGoto}
S_M = \mathcal{M}\int_\Sigma \sqrt{-\gamma},d\lambda^0 d\lambda^1 = \mathcal{M} \int_\Sigma \left[-\frac{1}{2}\Sigma_{\mu\nu}\Sigma^{\mu\nu}\right]^{1/2} d\lambda^0 d\lambda^1.
\end{equation}
Here, $\lambda^a = (\lambda^0, \lambda^1)$ are the worldsheet coordinates parameterizing the string trajectory \cite{Synge:1960ueh, Polchinski:1998rq}, $\mathcal{M}$ is the string tension (a positive constant with dimensions of energy per unit length), and $\gamma$ is the determinant of the induced metric on the string worldsheet:

\begin{equation}\label{induced_metric}
\gamma_{ab} = g_{\mu\nu}\frac{\partial x^\mu}{\partial\lambda^a}\frac{\partial x^\nu}{\partial\lambda^b}.
\end{equation}
The antisymmetric tensor $\Sigma^{\mu\nu}$ in Eq. (\ref{NambuGoto}) represents the bi-vector density of the string cloud \cite{Letelier:1979ej}, defined as:
\begin{equation}
\Sigma^{\mu\nu} = \epsilon^{ab}\frac{\partial x^\mu}{\partial\lambda^a}\frac{\partial x^\nu}{\partial\lambda^b},
\end{equation}
where $\epsilon^{ab}$ is the two-dimensional Levi-Civita symbol. This formalism provides a relativistic description of string-like matter distributions in curved spacetime \cite{Vilenkin:1981zs}.
The corresponding energy-momentum tensor takes the form \cite{Letelier:1979ej,Singh:2020nwo,Herscovich:2010vr}:
\begin{equation}\label{EMT}
T_{\mu\nu}^\text{cs} = \frac{\rho \Sigma_{\mu\sigma}\Sigma_{\nu}^{\;\sigma}}{\sqrt{-\gamma}},
\end{equation}
where the string bivector $\Sigma^{\mu\nu} = \epsilon^{ab}\partial_a x^\mu \partial_b x^\nu$ satisfies the conservation laws:
\begin{equation}\label{bivector_eqs}
\Sigma^{\mu\beta}\nabla_\mu\left[\frac{\Sigma_\beta^{\;\nu}}{(-\gamma)^{1/2}}\right] = 0, \quad \nabla_\mu(\rho\Sigma^{\mu\sigma})\Sigma_\sigma^{\;\nu} = 0.
\end{equation}The variation of the action \eqref{action} with metric $g_{\mu\nu}$ yields Einstein's equations with a nonzero cosmological constant are
\begin{equation}\label{EE}
R_{\mu\nu}-\frac{1}{2}g_{\mu\nu}R-3L^{-2} g_{\mu\nu}=T_{\mu\nu}^\mathrm{cs}.
\end{equation}

The original Letelier's  BH–CoS solution~\cite{Letelier:1979ej} considered only the electric-like component (\( \Sigma_{01} \)) with \( \gamma < 0 \), offering a framework to study black hole interactions with extended matter sources. \citet{Alencar:2025zyl} extended Letelier's model to include both electric (\(\Sigma_{01} \)) and magnetic-like (\(\Sigma_{23} \)) components, resulting in a spherically symmetric metric. In this study, we extend the setting work of \citet{Alencar:2025zyl}  by introducing a negative cosmological constant, which yields a BH-CoS-AdS metric
\begin{equation}\label{NCoS_metric}
ds^2 = -P(r)dt^2 + Q(r)dr^2 + S(r)(d\theta^2 + \sin^2\theta d\phi^2),
\end{equation}
with
\begin{equation}\label{metric_function}
P(r) =\frac{1}{Q(r)}\equiv 1 - \frac{2M}{r} + \frac{\lvert a\rvert c_0^2}{r^2}{}_2F_1\left(-\frac{1}{2},-\frac{1}{4},\frac{3}{4},-\frac{r^4}{c_0^4}\right)+\frac{r^2}{l^2},~~~~~~~~~S(r)=r^2.
\end{equation}
Here, $M$ means the BH mass, while $(\lvert a\rvert,~c_0)$ represent CoS. The solution reduces to the classical Letelier-AdS spacetime when $c_0 \to 0$ with $\lvert a\rvert = a$.
The thermodynamic quantities of the BH-AdS-CoS as functions of the event horizon radius  $r_+$ read
\begin{eqnarray}\label{thermo}
 M_+&=&\frac{r_+}{2}\left[1+r_+^2+\frac{\lvert a\rvert c_0^2}{2 r_+^2}{}_2F_1\left(-\frac{1}{2},-\frac{1}{4};\frac{3}{4};-\frac{r_+^4}{c_0^4}\right)\right],~~~~~~~~T_+=\frac{1}{4 \pi r_+}\left[1+3r_+^2-\frac{ \lvert a\rvert  \sqrt{r_+^4+c_0^4}}{r_+^2}\right]\nonumber\\
 S_+&=&\pi r_+^2, ~~~~~~~~~F_+=M_+-T_+S_+\equiv \frac{1}{4r_+}\left[r_+^2(1-r_+^2)+\lvert a\rvert\left(\sqrt{r_+^4+c_0^4}+2c_0 {}_2F_1\left(-\frac{1}{2},-\frac{1}{4};\frac{3}{4};-\frac{r_+^4}{c_0^4}\right)\right)\right].
 \end{eqnarray} All the quantities defined in Eq. \eqref{thermo} are dimensonless such that $r_+=r_+/l,~c_0=c_0/l,~M_+=M_+/l,~T_+=T_+l,~S_+=S_+/l^2, ~\text{and}~F_+=F_+/l$.

\begin{figure*}
  \centering
  \begin{tabular}{p{6cm} p{6cm} p{6cm}}
    \includegraphics[width=0.35\textwidth]{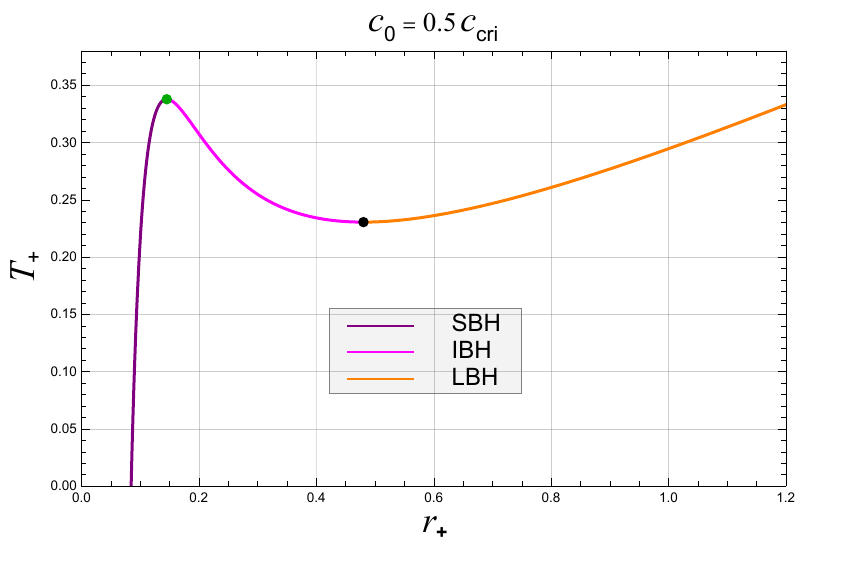}&
    \includegraphics[width=0.35\textwidth]{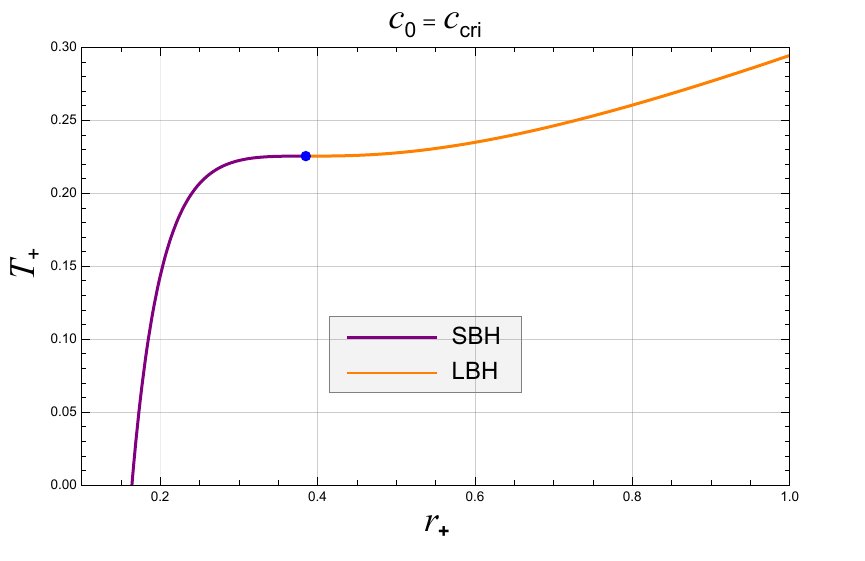}&
    \includegraphics[width=0.35\textwidth]{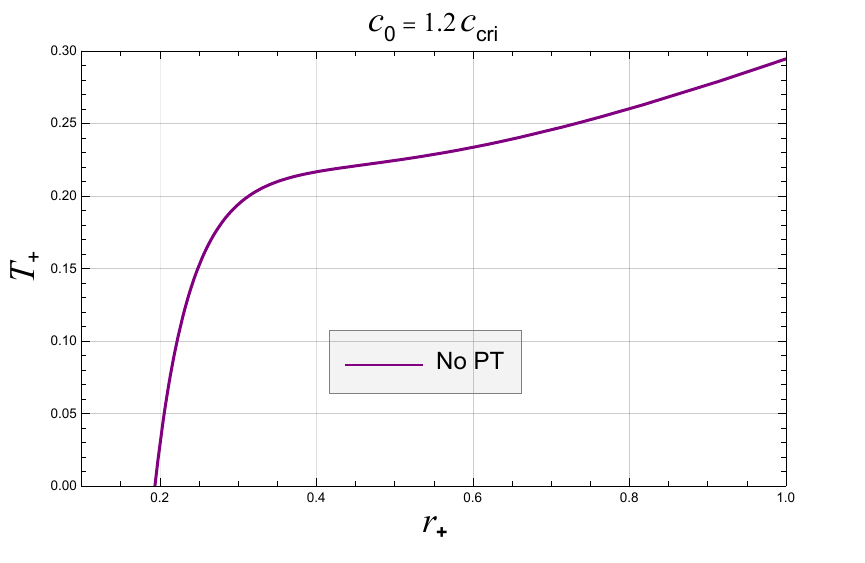}\\
    \includegraphics[width=0.35\textwidth]{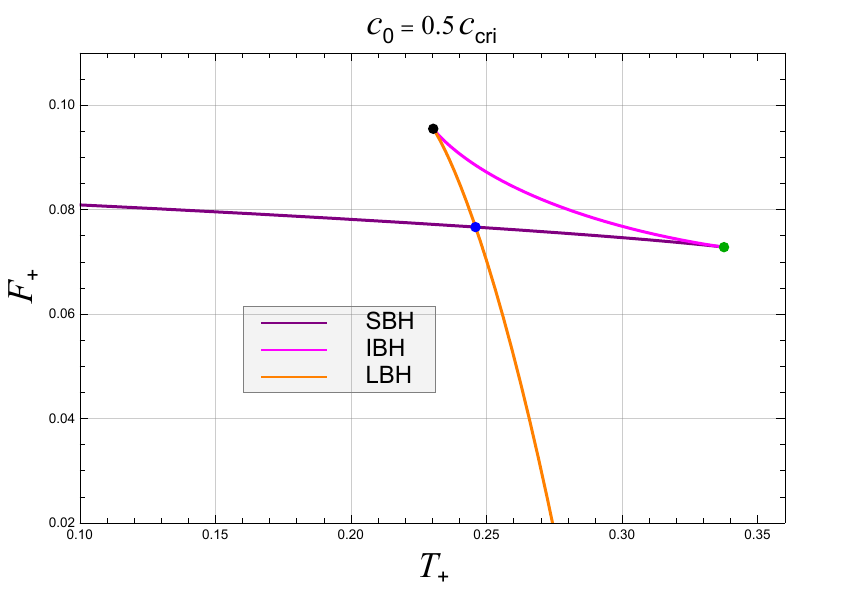}&
    \includegraphics[width=0.35\textwidth]{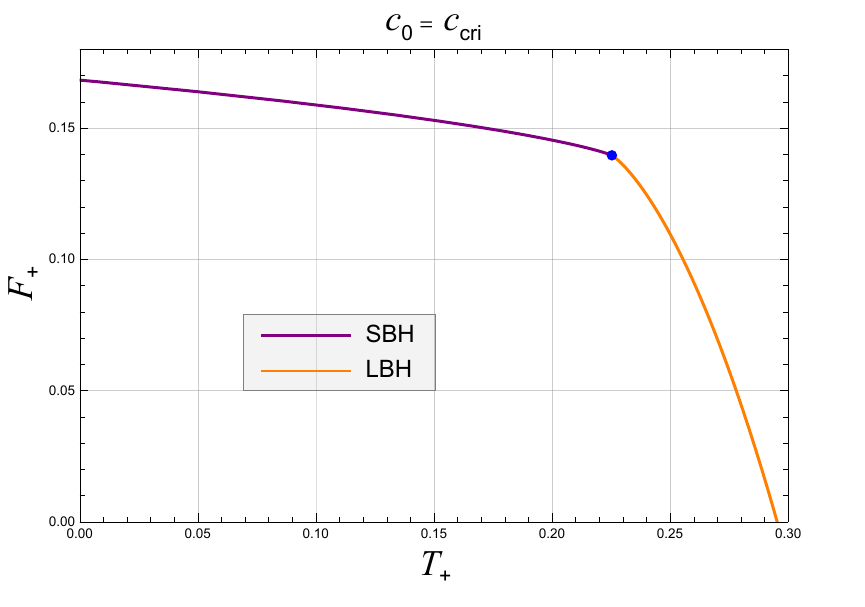}&
    \includegraphics[width=0.35\textwidth]{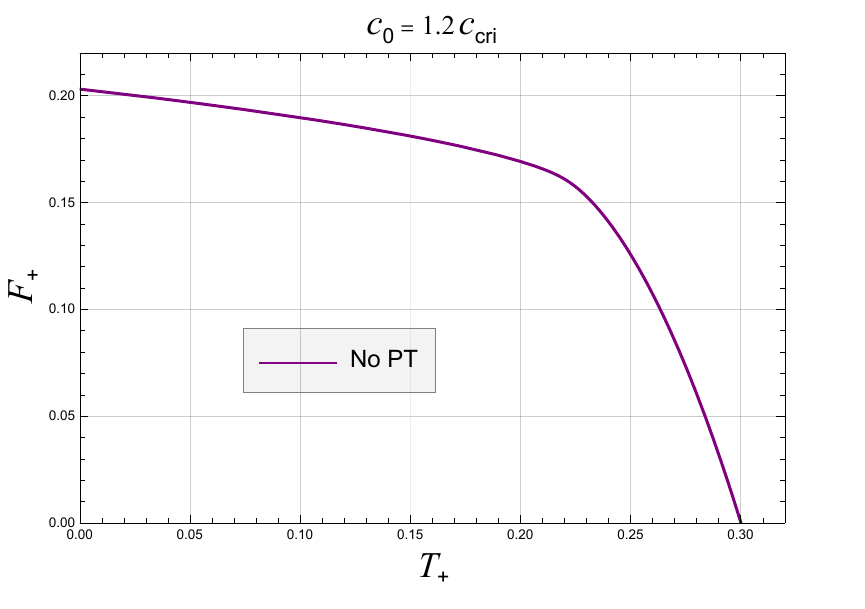}
\end{tabular}
  \caption{Temperature $T_+$ as function of $r_+$ (upper panel) and Free energy $F_+$ as a function of $T_+$ (lower panel)
    for BH-AdS-CoS with $\lvert a\rvert=0.3$. \textbf{Left:} $c_0<c_{\text{cri}}$. There
    coexist three BH solutions when $T_{1}~(\text{black dot})<T_+<T%
      _{2}~(\text{green dot})$, and a first-order phase transition between Small BH and Large BH occurs
    at $T_{p}~(\text{blue dot})$. \textbf{Middle:} $c_0>c_{\text{cri}}$. For $T_+<T_{\text{cri}}$ SBH and for $T_+>T_{\text{cri}}$ LBH and second-order phase transition at $T_{\text{cri}}$.  \textbf{Right:} $c_0>c_{\text{cri}}$. Only one
    BH solution exists, and hence no phase transition occurs.}%
  \label{pt}%
\end{figure*}

Using Eq.~\eqref{thermo}, we can express the horizon radius $r_{+}$ as a function of the temperature $T_+$. The multivalued nature of $r_{+}(T_{+})$ indicates the coexistence of multiple BH solutions for fixed parameters ($c_0$, $\lvert a\rvert$, $T_+$), corresponding to distinct phases in a canonical ensemble \cite{Kubiznak:2012wp, Chamblin:1999tk}. The critical point, characterized as an inflection point, satisfies
\begin{equation}\label{PT}
  \frac{\partial T_+}{\partial r_{+}} = 0, \quad 
  \frac{\partial^{2} T_+}{\partial r_{+}^{2}} = 0,
\end{equation}
yielding critical values $r_{\text{cri}}$, $c_{\text{cri}}$, and $T_{\text{cri}}$ whose analytical forms are obstructed by the complexity of Eq.~\eqref{PT} \cite{Chamblin:1999hg}. 

For analysing the phase structure of the considered system, we express the free energy $F_+$ in terms of $T_+$ by substituting $r_{+}(T_+)$ into $F_+$, and plot $T_+$ versus $r_+$ and $F_+$ versus $T_+$ for varying $c_0$ with $\lvert a\rvert = 0.3$ (Fig.~\ref{pt}). For $c_0 < c_{\text{cri}}$, three phases emerge: small BH (SBH), intermediate BH (IBH), and large BH (LBH)
coexisting within a temperature range $T_1 < T_+ < T_2$, while for $T_+<T_1$ and $T_+>T_2$ there exists only SBH and LBH, respectively. At $T_+=T_1$, IBH and LBH coexist, whereas at $T_+=T_2$, we observe the coexistence of SBH and IBH.  A first-order transition occurs at $T_p$ (blue dot, left panel), while at $c_0 = c_{\text{cri}}$, a second-order transition connects SBH and LBH at $T_{\text{cri}}$ (middle panel) \cite{Hawking:1982dh}. For $c_0 > c_{\text{cri}}$, only one stable phase exists (right panel). The upper and lower panels of Fig.~\ref{pt} demonstrate analogous behavior for $T_+(r_+)$ and $F_+(T_+)$, respectively \cite{Letelier:1979ej,Singh:2020xju}.

\section{Phase Transitions and Lyapunov Exponents of Particles}

\label{sec:particle}
The dynamics of test particles in BH spacetimes can encode vital information about the underlying geometry and thermodynamic properties of the background, including phase transitions \cite{Cardoso:2008bp, Konoplya:2017wot, Guo:2022kio}. The Lyapunov exponent, which describes the instability timescale of circular geodesics, has emerged as a strong diagnostic of BH criticality \cite{Chen:2025xqc, Lyu:2023sih}. This approach reveals a deep connection between geodesic instability and the nature of phase transitions in BH spacetimes.  We investigate the thermodynamic phase structure of a BH-AdS-CoS by analysing the  Lyapunov exponents associated with the unstable motion of the photons and massive particles \cite{Yang:2023hci}.
 
 Specifically, we concentrate on unstable circular geodesics that are defined by the Lagrangian on the equatorial hyperplane with $\theta=$ $\pi/2$
\begin{equation}
  2\mathcal{L}= g_{\mu\nu} \dot{x}^{\mu}\dot{x}^{\nu}= -P\left(  r\right)  \dot{t}+Q\left(  r\right)  \dot
  {r}^{2}+S(r)\dot{\phi}^{2}.
\end{equation}
Dots are employed to denote derivatives concerning the proper time. Using the definition of generalized momenta $$p_{\mu}=\frac{\partial\mathcal{L}}{\partial\dot{x}^{\mu}} = g_{\mu\nu}\dot{x}^{\nu},$$
the spherical symmetry of metric \eqref{NCoS_metric} leads to two conserved quantities $E$ (photon's energy) and $L$ (angular momentum), in the direction of $t$ and $\phi$ \cite{Chandrasekhar:1985kt}
\begin{eqnarray}
    P(r)\dot{t}=-E~~~~~~~~~~~~~~~\text{and}~~~~~~~~~~~~S(r) \dot{\phi}=L.
    \end{eqnarray}
The motion of massless particles is governed by the null geodesic equations. By employing the definition of the effective potential, denoted as $V(r) = \dot{r}^2$, we derive an expression for the effective potential of a massless particle within the context of BH-AdS-CoS as
\begin{eqnarray}\label{Veff}
  &&V\left(  r\right)  =\frac{1}{Q(r)}\left[  \frac{L^{2}}{S(r)}-\frac{E^2}{P(r)}+\delta\right]\equiv\nonumber\\&&  \delta-E^2+\frac{L^2(r-r_+)\left(1+r^2+r_+^2+rr_+\right)}{r^3}+\frac{\lvert a\rvert c_0^2L^2}{r^4r_+}\left[r_+{}_2F_1\left(-\frac{1}{2},-\frac{1}{4},\frac{3}{4},-\frac{r^4}{c_0^4}\right)-r{}_2F_1\left(-\frac{1}{2},-\frac{1}{4},\frac{3}{4},-\frac{r_+^4}{c_0^4}\right)\right] ,
\end{eqnarray}
where $\delta=1$ and $0$ correspond to massive and massless particles, respectively. We initially determine the position of the unstable circular orbit for the particle, which is governed by the conditions
\begin{equation}\label{conditions}
  V_r(r)  =0,\ V_{rr}\left(  r\right)  <0,
\end{equation} 
where $V_r(r)=\partial V(r)/\partial r$ and $V_{rr}(r)=\partial^2 V(r)/\partial r^2$. The Lyapunov exponent $\lambda$ can be expressed as \cite{Gogoi:2024akv, Guo:2022kio, Kumara:2024obd, Cardoso:2008bp}
\begin{equation}\label{Lyapunov}
    \lambda=\sqrt{-\frac{V_{rr}(r_\text{o})}{2\dot{t}^2}}
\end{equation}
with $r_{\text{0}}$ being the circular orbit radius.

It is crucial to note that we will explore the unstable circular geodesics and the relationship between the Lyapunov exponents associated with these circular orbits and the phase structure of the system. Motions close to unstable circular orbits are extremely sensitive to initial conditions, as seen by the exponential increase in perturbation that occurs when unstable circular motions are disturbed. Thus, chaotic motions in BHs and unstable circular motions are closely related, and understanding these motions could be very helpful in understanding the characteristics of chaos. Lyapunov exponents, which quantify the rate of divergence of nearby trajectories, emerge as a key diagnostic tool. Specifically, Maldacena, Shenker, and Stanford proposed in Ref.~\cite{Maldacena:2015waa} that the Lyapunov exponent $\lambda$ in thermal quantum systems is bounded above by $\lambda \leq 2\pi T$, where $T$ is the temperature—a conjecture now widely referred to as the \textit{chaos bound}. This bound has profound implications in holography and quantum gravity. However, it was discovered in \cite{Lei:2021koj} that unstable circular motions of charged particles close to the charged BH do not follow the bound defined in \cite{Maldacena:2015waa}. Furthermore, unstable circular null geodesics—collectively forming the so-called photon sphere just outside the event horizon—play a crucial role in BH imaging and observational phenomena such as gravitational lensing and shadow formation. The dynamics of photons near this region encode rich information about the BH geometry, and analysing their instability through Lyapunov exponents sheds light on the phase structure and connects theory to observational signatures.

\subsection{Massless Particles}

For massless particles, i.e., photons, the circular orbit radius can be determined by setting $\delta = 0$ in the effective potential expression~\eqref{Veff}, and applying the conditions for unstable circular geodesics as given in Eq.~\eqref{conditions}. This yields the relation between the conserved quantities $L$ and $E$ as
\begin{equation}
    \frac{L}{E} = \frac{r_0}{\sqrt{P(r_0)}}.
\end{equation} Substituting this relation into Eq.~\eqref{Lyapunov}, the time component of the geodesic motion becomes
\begin{equation}
    \dot{t} = \frac{L}{r_0 \sqrt{P(r_0)}},
\end{equation}
and the Lyapunov exponent, which quantifies the instability of the orbit, simplifies to
\begin{equation}
    \lambda = \sqrt{ -\frac{r_0^2 P(r_0)}{2L^2} V_{rr}(r_0) }.
\end{equation}

\begin{figure}[h]
  \centering
  \includegraphics[width=8cm]{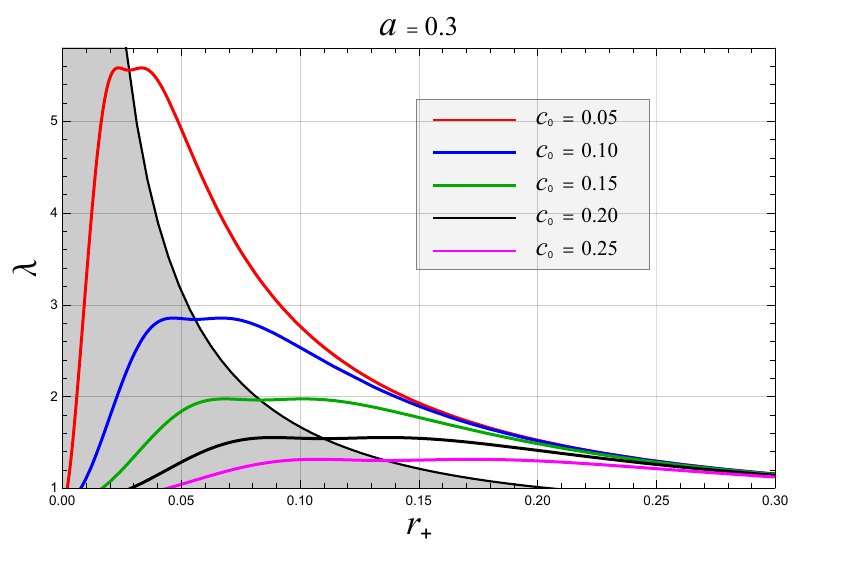}
  \caption{The Lyapunov exponent $\lambda$ vs. horizon radius $r_{+}$ for unstable null circular geodesics, plotted for various values of the cloud-of-strings parameter $c_0$ with fixed $a=0.3$. The gray region denotes the non-physical regime where the Hawking temperature becomes negative.}
  \label{density1}
\end{figure}

In Fig.~\ref{density1}, we illustrate the numerical behaviour of $\lambda$ as a function of the event horizon radius $r_+$ for several values of the string cloud parameter $c_0$, with $a = 0.3$ fixed. The shaded region corresponds to non-physical BHs with negative Hawking temperature ($T_+ < 0$). The plots show that for smaller horizon radii, the Lyapunov exponent $\lambda$ increases as $c_0$ decreases. However, for sufficiently large $r_+$, the values of $\lambda$ for different $c_0$ converge, indicating a universal dynamical behaviour in this regime. Furthermore, in the asymptotic limit $r_+ \to \infty$ (or equivalently $c_0 \to \infty$), the Lyapunov exponent approaches unity, $\lambda \to 1$, consistent with anticipations for highly stable LBH configurations. This behaviour of $\lambda$ highlights its sensitivity to the BH's microstructure and thermodynamic phase, especially in the near-horizon. 

\begin{figure*}
   \centering
  \begin{tabular}{p{6cm} p{6cm} p{6cm}}
    \includegraphics[width=0.35\textwidth]{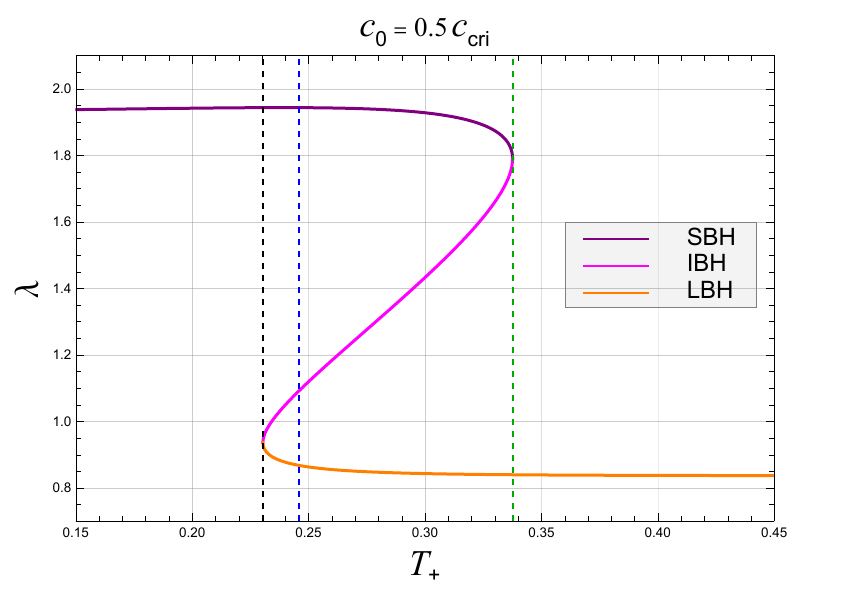}& 
    \includegraphics[width=0.35\textwidth]{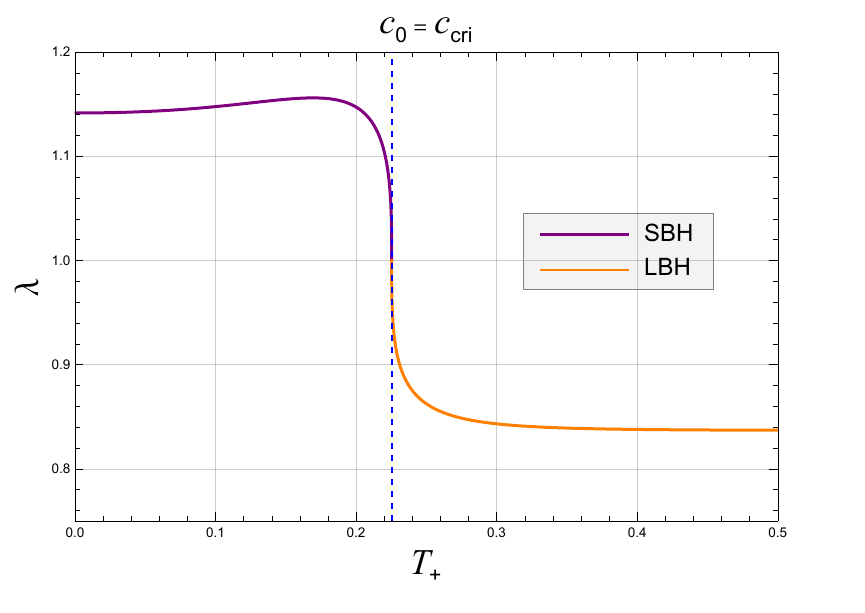}&
    \includegraphics[width=0.35\textwidth]{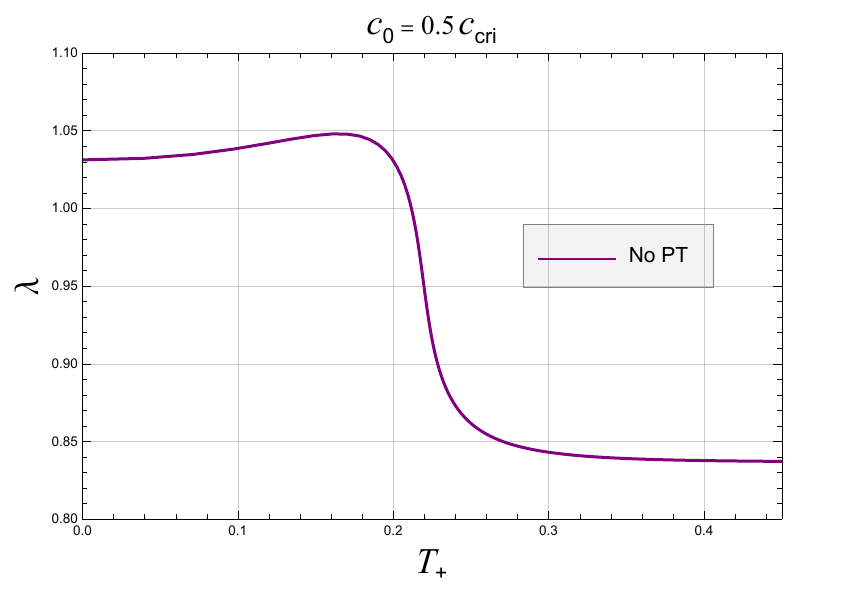} 
  \end{tabular}
  \caption{The Lyapunov exponents $\lambda$ of photons as a function of temperature $T_+$ for $c_0<c_{\text{cri}}$ (\textbf{Left Column}), $c_0=c_{\text{cri}}$ (\textbf{Middle Column}), and $c_0>c_{\text{cri}}$ (\textbf{Right Column}). Three BH solutions, SBH, IBH, and LBH, coexist for $T_{1} (\text{dashed black vertical line})<T_+<T_{2}$ (dashed green vertical line). The phase shift from SBH to LBH happens at $T_+=T_{p}$ (dashed blue vertical line). For $c_0<c_{\text{cri}}$, $\lambda$ is a multivalued function of $T_+$, with three branches corresponding to three BH phases, respectively. When $c_0>c_{\text{cri}}$, $\lambda$ is single-valued, signifying single phase of BH and $c_0=c_{\text{cri}}$ signifies the second-order phase transition between SBH and LBH.}%
  \label{fig:lambdaTN}%
\end{figure*}

 The Lyapunov exponent $\lambda$ can be parametrized as a function of the BH temperature $T_+$ by inverting the Hawking temperature relation of Eq.~\eqref{thermo} to obtain $r_+(T_+)$. Fig.~\ref{fig:lambdaTN} illustrates $\lambda(T_+)$ for various values of the string cloud parameter $c_0$, thereby exposing distinct thermodynamic behaviours.
In the left panel, for $c_0 = 0.5\,c_{\rm cri}$, three BH phases coexist over the temperature interval $T_1 < T_+ < T_2$: small (SBH), intermediate (IBH), and large (LBH). In this regime, the function $\lambda(T_+)$ becomes multivalued, with three separate branches precisely tracking the respective phases—demonstrating the conjectured correspondence between Lyapunov exponents and BH thermodynamic phases \cite{Guo:2022kio}. At $T_+=T_1$, $\lambda$ of IBH merges with that of LBH, whereas at $T_+=T_2$, IBH and SBH have the same $\lambda$.

At $T_+ = T_p$, a first-order transition transpires between SBH and LBH, displaying as a discontinuity in $\lambda$, which thus acts as a dynamical order parameter. In the SBH branch, $\lambda$ initially rises to a local maximum before decreasing as $T_+$ approaches $T_2$. By contrast, in the IBH branch $\lambda$ increases monotonically, while in the LBH branch, it decreases as $T_+$ increases from $T_1$. All branches asymptotically approach $\lambda \to 1$ as $T_+ \to \infty$, consistent with general high-temperature stability trends \cite{Shukla:2024tkw, Chen:2025xqc}.

The system exhibits a single thermodynamic phase across all temperatures in the right panel ($c_0 = 1.2\,c_{\rm cri}$). Consequently, $\lambda(T_+)$ is single-valued and smooth, rising to a peak then declining toward unity with increasing $T_+$. At the critical value $c_0 = c_{\rm cri}$ (middle panel), the system undergoes a second-order transition: SBH and LBH branches merge continuously at $T_+ = T_{\rm cri}$. Here $\lambda(T_+)$ is continuous but exhibits non-analyticity at the critical temperature, indicating critical behaviour \cite{Shukla:2024tkw, Kumara:2024obd}.  These observations affirm that the Lyapunov exponent $\lambda$ serves as an effective probe of BH phase structure. Its multivalued nature in first‑order transitions, the continuity with an inflection point at criticality, and its asymptotic limit confirm its role as a dynamical indicator of thermodynamic phases \cite{Guo:2022kio, Kumara:2024obd, Chen:2025xqc}.

\begin{figure}[t]
  \begin{center}
    \includegraphics[width=0.48\textwidth]{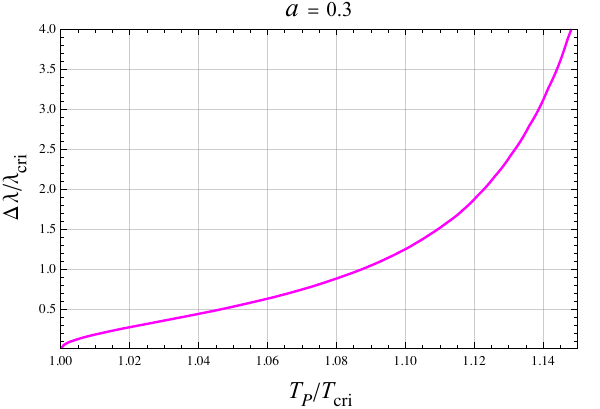}\hspace{1em}
    \includegraphics[width=0.48\textwidth]{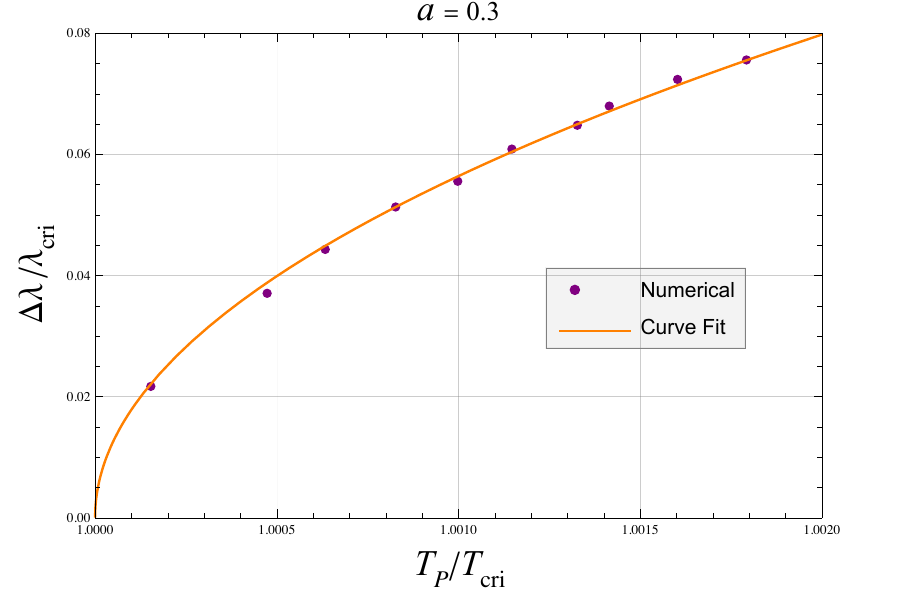}
  \end{center}
  \caption{Rescaled discontinuity in the Lyapunov exponent $\Delta
      \lambda/\lambda_{\text{cri}}$ during the phase transition as a function of the reduced phase transition temperature $t\equiv{T}_{p}/{T}_{cri}$ near the
    critical temperature $t=1$.}%
  \label{dlN}%
\end{figure}

A fascinating aspect of the phase transition in the BH-AdS-CoS is the existence of a discontinuity in the Lyapunov exponent, which serves as a diagnostic for dynamical instability, which is defined as $\Delta \lambda = \lambda_S - \lambda_L$, where $\lambda_S$ and $\lambda_L$ denote the Lyapunov exponents corresponding to SBH and LBH phases, respectively, evaluated at the first-order transition temperature $T_+ = T_P$. At the critical point, where $T_+ = T_{\text{cri}}$, the system experiences a second-order phase transition and the two branches coincide, i.e., $\lambda_S = \lambda_L = \lambda_{\text{cri}}$, implying that $\Delta \lambda = 0$.

The Fig.~\ref{dlN} (left panel) depicts the normalised discontinuity $\Delta \lambda/\lambda_{\text{cri}}$ as a function of the reduced transition temperature $t \equiv T_P/T_{\text{cri}}$, with $t=1$ marking the critical point. The plot reveals that during the first-order transition from SBH to LBH, the Lyapunov exponent exhibits a finite jump, thus yielding a non-zero $\Delta \lambda$. It shows $\Delta \lambda$ as an effective dynamical order parameter, consistent with its proposed role in recent literature \cite{Guo:2022kio, Yang:2023hci, Zhao:2018wkl}.

To further explore the critical behaviour, we expand $\Delta \lambda$ near the critical point ($t \approx 1$), obtaining the scaling relation:
\begin{equation}
  \frac{\Delta\lambda}{\lambda_\text{cri}} \sim 1.78374\sqrt{t - 1},
\end{equation}
which reveals that the critical exponent associated with $\Delta \lambda$ is $\beta = 1/2$. This value precisely matches the critical exponent predicted by mean-field theory for conventional thermodynamic systems such as the van der Waals fluid \cite{Kubiznak:2012wp, Chamblin:1999tk}. Fig.~\ref{dlN} (right panel) further shows the behaviour of $\Delta \lambda$ in the vicinity of the critical point. The numerical data, represented by purple dots, aligns excellently with the analytical fit shown by the orange curve, reinforcing the universality of the critical exponent $\beta = 1/2$. 

\subsection{Massive Particles}

In the BH-AdS-CoS background, massive test particles can admit both stable and unstable circular geodesics, contingent upon the value of the event horizon radius $r_{+}$ and other spacetime parameters. The unstable time-like circular orbits are particularly interesting, as they play a critical role in understanding BH dynamics and thermodynamic phase transitions. These orbits are especially relevant due to their connection with dynamical instability and their relation to the proposed universal upper bound on Lyapunov exponents, $\lambda \leq 2\pi T$, in the context of holography and quantum chaos \cite{Hashimoto:2016dfz, Maldacena:2015waa, Zhao:2018wkl}.

\begin{figure}[pt]
  \centering
  \includegraphics[width=0.5\textwidth]{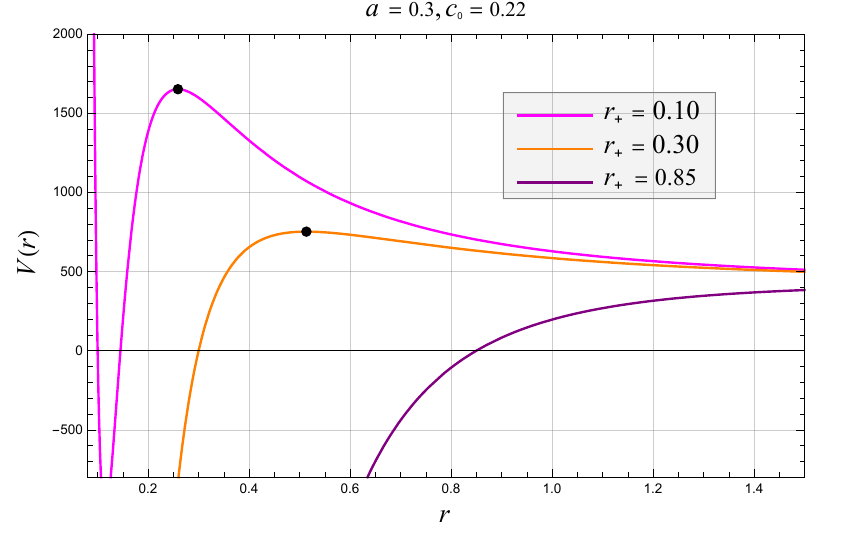}
  \caption{The effective potential for massive particles with angular momentum $L=20l$ in the BH-AdS-CoS spacetime for parameters $c_0=0.22$, $a=0.3$, and horizon radii $r_{+}=0.10$, $0.30$, and $0.85$. The black dots indicate the local maxima of the effective potential, corresponding to unstable time-like circular orbits. Notably, the effective potential exhibits no maximum for $\tilde{r}_{+}=0.85$, implying the absence of such unstable geodesics.}
  \label{Veff2}
\end{figure}

Our analysis focuses on computing the Lyapunov exponent $\lambda$ corresponding to such unstable time-like circular orbits, assuming a fixed angular momentum $L$. Fig.~\ref{Veff2} illustrates the behaviour of the effective potential $V(r)$ for massive particles in the BH-AdS-CoS spacetime with $c_0=0.22$ and $a=0.3$, plotted for various values of $r_+$. For smaller horizon radii, the potential exhibits a local maximum, indicating the existence of unstable circular orbits (the pink and orange curves in Fig. \ref{Veff2}). However, as the horizon radius increases, the peak of the potential vanishes. For instance, the potential fails to develop a maximum for $r_{+} = 0.85$, indicating the absence of any unstable time-like circular orbit, as shown by the purple curve in Fig. \ref{Veff2}. 

\begin{figure*}  \centering
  \includegraphics[width=0.5\textwidth]{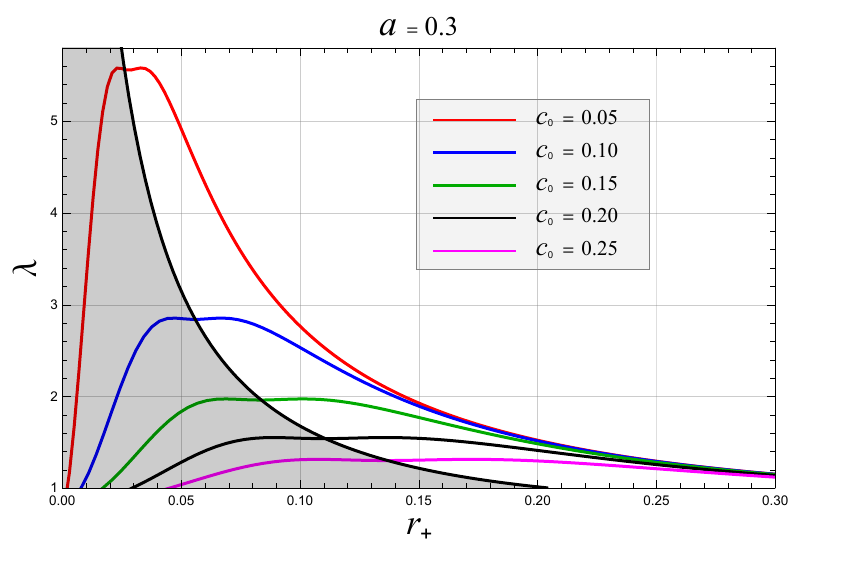} 
  \caption{The Lyapunov exponent $\lambda$ as function $r_{+}$ for unstable time-like circular geodesics for different values of $c_0$ with $a=0.3$. The gray region signifies the non-physical region (negative Hawking temperature).}%
  \label{density2}%
\end{figure*}

By imposing the conditions for timelike geodesics $\delta=1$, $V(r_{\text{0}})=0$, and $V_{r}(r_{\text{0}})=0$, we obtain the following relation for energy and momentum of massive particles 
\begin{equation}\label{LE}
    L^2=\frac{r_{\text{o}}^3P_r(r_{\text{0}})}{2P(r_{\text{o}})-r_{\text{o}}P_r(r_{\text{o}})},~~~~~~~~~~\text{and}~~~~~~~~~~E^2=\frac{2P^2(r_{\text{0}})}{2P(r_{\text{o}})-r_{\text{o}}P_r(r_{\text{0}})},
\end{equation} which leads to
\begin{equation}
    \dot{t}=\frac{1}{\sqrt{P(r_{\text{o}})-\frac{1}{2}r_{\text{o}}P_r(r_{\text{o}})}}.
\end{equation}
Hence, the Lyapunov exponent for time-like circular geodesics can be described as \cite{Cardoso:2008bp}
\begin{equation}
  \lambda = \frac{1}{2} \sqrt{ \left[ 2P(r_{\text{o}}) - r_{\text{o}} P_r(r_{\text{o}}) \right] V_{rr}(r_{\text{o}}) }, 
  \label{eq:lamdaM}
\end{equation}
where $r_{\text{o}}$ represents the radius of the unstable circular orbit, $P(r)$ is the metric function, and $V_{rr}(r)$ is the second-order derivative of the effective potential with respect to the radial coordinate. This expression captures the instability timescale associated with time-like circular geodesics near the BH and can be recast in terms of the model parameters $c_0$, $r_+$, and $L$ using Eq.~\eqref{LE}. The behaviour of the Lyapunov exponent $\lambda$ for unstable time-like geodesics as a function of the event horizon radius $r_{+}$ is shown in Fig.~\ref{density2}, for fixed angular momentum $L=20l$ and  CoS parameter $a=0.3$, across several values of $c_0$. Notably, $\lambda$ exhibits similar trends to the null geodesic case at small $r_+$, indicating that both types of orbits experience comparable dynamical instability in this regime.

However, a significant departure arises at larger $r_+$: the Lyapunov exponent for time-like geodesics vanishes beyond a critical horizon radius, as confirmed by the absence of local maxima in the effective potential, demonstrated by the purple curve in Fig.~\ref{Veff2}. This signals the disappearance of unstable time-like circular orbits and implies a dynamically stable regime where test particles no longer exhibit exponential sensitivity to initial conditions. Such behaviour is characteristic of the LBH phase in the presence of a CoS and contrasts with the massless case, where $\lambda$ asymptotically approaches a non-zero constant at large $r_+$ values \cite{Konoplya:2017wot, Maldacena:2015waa, Yang:2023hci}.

\begin{figure*}
   \centering
  \begin{tabular}{p{6cm} p{6cm} p{6cm}}
    \includegraphics[width=0.35\textwidth]{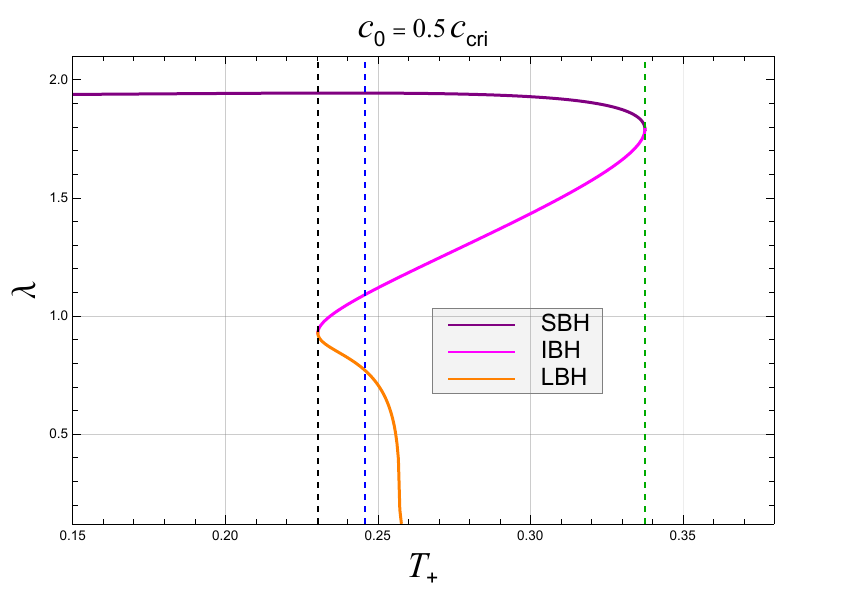}& 
    \includegraphics[width=0.35\textwidth]{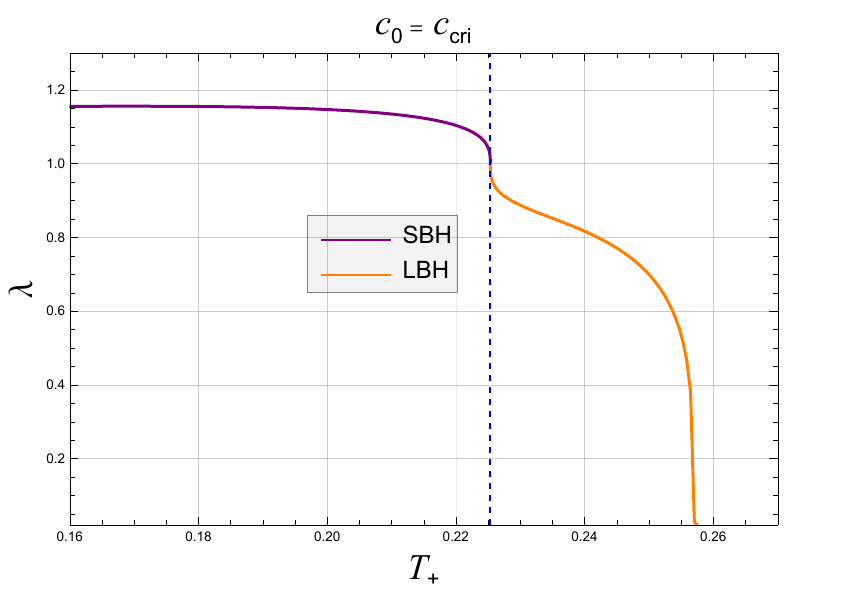}&
    \includegraphics[width=0.35\textwidth]{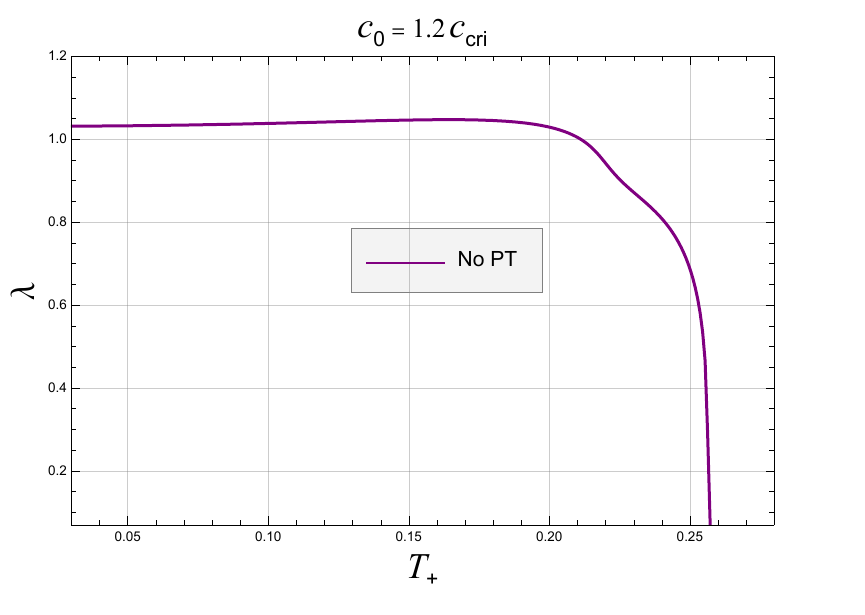} 
  \end{tabular}
  \caption{The Lyapunov exponents $\lambda$ of massive particles as a function of temperature $T_+$ for $c_0<c_{\text{cri}}$ (\textbf{Left Column}), $c_0=c_{\text{cri}}$ (\textbf{Middle Column}), and $c_0>c_{\text{cri}}$ (\textbf{Right Column}). Three BH solutions, SBH, IBH, and LBH, coexist for $T_{1} (\text{dashed black vertical line})<T_+<T_{2}$ (dashed green vertical line). The phase shift from SBH to LBH happens at $T_+=T_{p}$ (dashed blue vertical line). For $c_0<c_{\text{cri}}$, $\lambda$ is a multivalued function of $T_+$, with three branches corresponding to three BH phases, respectively. When $c_0>c_{\text{cri}}$, $\lambda$ is single-valued, signifying single phase of BH and $c_0=c_{\text{cri}}$ signifies the second-order phase transition between SBH and LBH.}%
  \label{fig:lambdaT}%
\end{figure*}

To understand the phase structure of BH-AdS-CoS through the lens of Lyapunov exponents of unstable circular orbits of massive particles, we presented the numerical results of $\lambda$ as a function of $T_+$ with $a=0.3$ and $L=20$ for different values of $c_0$ in Fig. \ref{fig:lambdaT}. We find that when the CoS parameter $c_0$ is less than its critical value $c_{\text{cri}}$, the Lyapunov exponent $\lambda$ becomes a multivalued function of $T_+$, exhibiting three distinct thermal branches. In the small stable BH-AdS-CoS phases, $\lambda$ firstly increases and then decreases with increasing $T_+$, whereas in the intermediate (large), thermodynamically unstable (stable) phase, it increases (decreases) with $T_+$. This branching behaviour strongly supports the interpretation of $\lambda$ as a dynamical order parameter for the phase transition \cite{Guo:2022kio, Yang:2023hci, Kumara:2024obd}.

In contrast, for $c_0>c_{\text{cri}}$, the Lyapunov exponent $\lambda$ transitions to a single-valued, monotonic function that decreases continuously with $T_+$. This shift from multivalued to single-valued behaviour of $\lambda$ effectively signals the presence of a critical point at $c_0=c_{\text{cri}}$. Thus marks the critical point and signals the onset of second-order phase transition behaviour, consistent with mean-field theory expectations \cite{Wei:2022dzw, Lyu:2023sih}. 

The qualitative features of the Lyapunov exponent for massive particles mirror those observed in the massless case, particularly in the SBH regime, where increasing $c_0$ suppresses the instability. However, key differences arise in the LBH phase: for massive particles, the Lyapunov exponent $\lambda$ tends to zero at large $T_+$, corresponding to a specific $r_+$ beyond which unstable circular orbits cease to exist. This phenomenon is evident from Fig.~\ref{fig:lambdaT}, where the disappearance of $\lambda$ occurs near the same value of $r_+$ for all values of $c_0$. This reflects the transition from chaotic to monotonic motion, as the effective potential no longer supports local maxima required for instability. 
Unlike massless particles, for which $\lambda$ asymptotically approaches a constant, massive particle orbits become stable and $\lambda$ vanishes, signalling a qualitatively different dynamical regime. 

\begin{figure}[t]
  \begin{center}
    \includegraphics[width=0.48\textwidth]{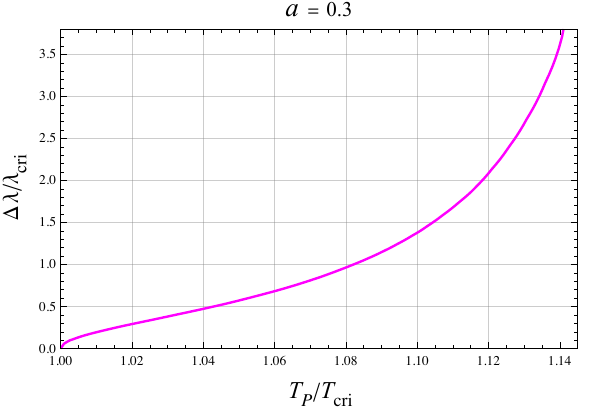}\hspace{1em}
    \includegraphics[width=0.48\textwidth]{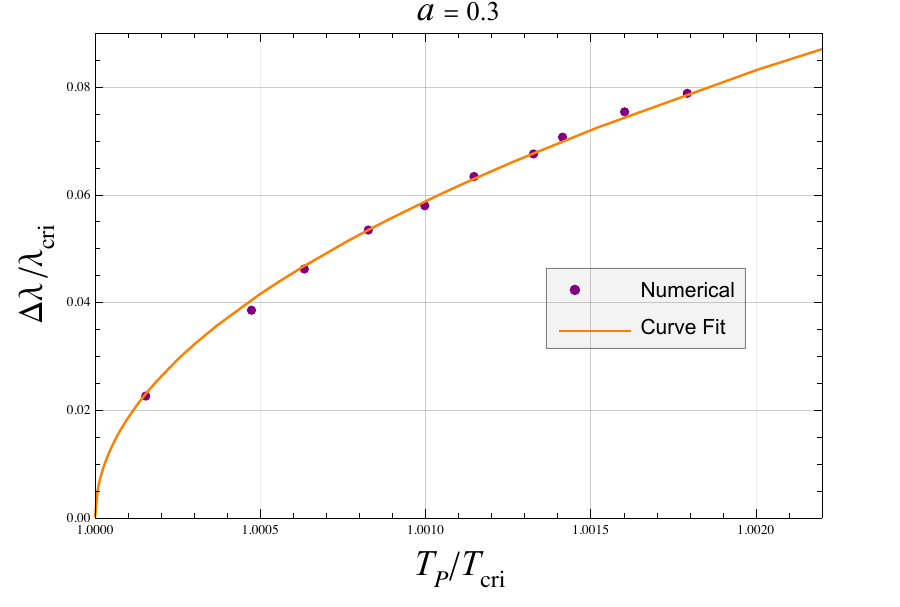}
  \end{center}
  \caption{Rescaled discontinuity in the Lyapunov exponent $\Delta
      \lambda/\lambda_{\text{cri}}$ during the phase transition as a function of the reduced phase transition temperature $t\equiv{T}_{p}/{T}_{cri}$ near the
    critical temperature $t=1$.}%
  \label{dl}%
\end{figure}

We now explore the Lyapunov exponent difference, $\Delta\lambda$, as a function of the reduced phase transition temperature for different values of the dimensionless CoS parameter, $c_0/c_{\text{cri}}$. Fig.~\ref{dl} describes the behaviour of the $\Delta\lambda/\lambda_{\text{cri}}$, plotted against the ratio $T_P/T_{\text{cri}}$. The figure shows that $\Delta\lambda$ remains finite and nonzero across the first-order phase transition, confirming the discontinuous nature of $\lambda$ during the transition between small and large black hole phases. As $T_P$ approaches the critical temperature $T_{\text{cri}}$, the discontinuity $\Delta\lambda$ reduces progressively and vanishes precisely at the critical point $t = T_P/T_{\text{cri}} = 1$, where $\Delta\lambda/\lambda_{\text{cri}} = 0$. This behaviour reflects the critical dynamics and supports the interpretation of $\Delta\lambda$ as an effective order parameter for the phase transition~\cite{Guo:2022kio, Yang:2023hci, Zhao:2018wkl}.
The right panel of Fig.~\ref{dl} zooms into the immediate vicinity of the critical point, highlighting the scaling behaviour of $\Delta\lambda$ near $t = 1$. The numerical data characterised by purple dots are well fitted by a square-root dependence, described analytically by
\begin{equation}
  \frac{\Delta\lambda}{\lambda_{\text{cri}}} \sim 1.85854\sqrt{t - 1},
\end{equation}
which confirms that the critical exponent associated with $\Delta\lambda$ is $1/2$. This value aligns with the universality class predicted by mean-field theory, further corroborating earlier results that suggest Lyapunov exponents can serve as dynamical order parameters near black hole criticality~\cite{Kumara:2024obd, Wei:2022dzw, Chen:2025xqc}.


\section{Conclusions} \label{sec:conclusion}
Understanding the thermodynamics of BHs remains one of the deepest challenges in gravitational physics, with far-reaching implications for both classical gravity and quantum theories of spacetime. In particular, recognising dynamical signatures of BH phase transitions has appeared as an active area of research, offering new perspectives on the interplay between geometry, matter fields, and statistical mechanics. The Lyapunov exponent, which was initially used as a gauge of chaos and orbital instability, is one of these dynamical indicators that has shown remarkable efficacy in examining the near-horizon structure and critical phenomena of black holes. Motivated by these developments, we examined the connection between BH thermodynamics and particle dynamics in the setting of BH-AdS-CoS and its relationship to the chaotic dynamics of test particles using Lyapunov exponents. 

Below, we summarise the key outcomes and insights of our investigation. A rich phase structure with discrete small, intermediate, and large BH phases, separated by first-order and second-order phase transitions that closely resemble the liquid-gas transition in van der Waals fluids, is revealed by our investigation of the thermodynamic properties of BH-AdS-CoS.  The critical behaviour was illustrated through careful examination of the temperature $T_+$ as a function of horizon radius $r_+$ and the free energy $F_+$ as a function of temperature, which apparently showed the characteristic swallowtail structure indicative of first-order transitions and the critical point marking a second-order transition.  The critical behaviour was demonstrated through the characteristic swallowtail structure in free energy $F_+(T_+)$ and multivalued behaviour in the temperature–horizon radius relation $T_+(r_+)$, with the critical point marking a second-order transition~\cite{Kubiznak:2012wp, Chamblin:1999tk}.

The central result of our work establishes the Lyapunov exponent $\lambda$ as a sensitive dynamical probe of these thermodynamic phase transitions. We derived exact analytical expressions for $\lambda$ for massless particles following null geodesics and demonstrated its multivalued behaviour in the regime where multiple BH phases coexist. Specifically, we discovered that $\lambda$ assumes distinct branches corresponding to the small, intermediate, and large BH phases, with the small BH branch establishing an initial increase to a maximum followed by a decrease with temperature. In contrast, the large BH branch (intermediate BH) monotonically decreases (increases) with temperature. Most significantly, close to the critical point, the discontinuity $\Delta\lambda$ in $\lambda$ throughout the first-order transition obeys a scaling law, 
\begin{equation} \frac{\Delta\lambda}{\lambda_\text{cri}} \sim 1.78374 \sqrt{t - 1}, \quad t \equiv \frac{T_p}{T_\text{cri}} \to 1 
\end{equation}
establishing a universal mean-field critical exponent $\beta = 1/2$, consistent with other works~\cite{Guo:2022kio,Yang:2023hci,Zhao:2018wkl}. This scaling behaviour persisted remarkably close to the critical point, as evidenced by our numerical results in Fig.~\ref{dlN}, where the purple data points closely follow the predicted square-root dependence.

A similarly rich structure was observed for massive particles following timelike geodesics. In this case, unstable circular orbits only exist below a critical horizon radius (the pink and orange curves in Fig.~\ref{Veff2}), beyond which the Lyapunov exponent $\lambda$ vanishes—signalling the absence of geodesic instability (the purple curve in Fig.~\ref{Veff2}). Notwithstanding this difference, the Lyapunov exponent continues to exhibit multivalued behavior in the vicinity of the phase transition and follows a comparable scaling law, 
\begin{equation} \frac{\Delta\lambda}{\lambda_\text{cri}} \sim 1.85854 \sqrt{t - 1}, \end{equation}, 
thereby confirming its function as a dynamical order parameter.
The slight variation in the prefactor between the massless and massive cases suggests a subtle dependence of critical dynamics on particle mass~\cite{Chen:2025xqc, Blaga:2023exi}.

The striking agreement between thermodynamic features and the behaviour of Lyapunov exponents—across both null and timelike geodesics—strongly supports the conjecture that chaotic dynamics encode key information about the thermodynamic phase structure of BHs. Our results not only establish the Lyapunov exponent as a powerful dynamical diagnostic of criticality, but also open avenues for its application in more general settings, including rotating or charged BHs, and those embedded in modified gravity theories or surrounded by exotic matter fields~\cite{Wei:2022dzw, Kumara:2024obd}.

Furthermore, the precise correspondence between dynamical instability (via $\lambda$) and thermodynamic transitions may have profound implications in the context of gauge/gravity duality, where Lyapunov exponents are intimately linked to quantum chaos and thermalisation in the dual field theory~\cite{Maldacena:2015waa, Hashimoto:2016wme}. In this light, our study reinforces the idea that Lyapunov exponents can serve as a bridge connecting gravitational dynamics in the bulk to critical phenomena on the boundary. Our results motivate future investigations into rotating or charged BH-AdS-CoS spacetimes, where frame dragging or electromagnetic interactions may yield richer dynamical structures.

\begin{acknowledgments}

This work is supported by the Zhejiang Provincial Natural Science Foundation of China under Grants No.~LR21A050001 and No.~LY20A050002, the National Natural Science Foundation of China under Grants No.~12275238 and No. ~W2433018, the National Key Research and Development Program of China under Grant No. 2020YFC2201503, and the Fundamental Research Funds for the Provincial Universities of Zhejiang in China under Grant No.~RF-A2019015. SGG acknowledges support from ANRF under the project CRG/2021/005771.
 
\end{acknowledgments}

\appendix

\bibliography{reference.bib}
\bibliographystyle{apsrev4-1}
\end{document}